\documentclass[12pt]{article}
\usepackage{graphicx}

\catcode`\@=11
\def\marginnote#1{}

\relax

\def\beq{\begin{equation}}
\def\eeq{\end{equation}}
\def\bea{\begin{eqnarray}}
\def\eea{\end{eqnarray}}
\def\beaa{\begin{array}}
\def\eeaa{\end{array}}

\begin{document}

\title{\hspace*{-13mm}Character decomposition of Potts model partition functions. \hspace*{-13mm} \\
       I. Cyclic geometry}

\author{
  {\small Jean-Fran\c{c}ois Richard${}^{1,2}$ and
          Jesper Lykke Jacobsen${}^{1,3}$} \\[1mm]
  {\small\it ${}^1$Laboratoire de Physique Th\'eorique
  et Mod\`eles Statistiques}                             \\[-0.2cm]
  {\small\it Universit\'e Paris-Sud, B\^at.~100,
             91405 Orsay, France}                        \\[1mm]
  {\small\it ${}^2$Laboratoire de Physique Th\'eorique
  et Hautes Energies}                                    \\[-0.2cm]
  {\small\it Universit\'e Paris VI,
             Bo\^{\i}te 126, Tour 24, 5${}^{\mbox{\`eme}}$ {\'e}tage} \\[-0.2cm]
  {\small\it 4 place Jussieu, 75252 Paris cedex 05, France} \\[1mm]
  {\small\it ${}^3$Service de Physique Th\'eorique}      \\[-0.2cm]
  {\small\it CEA Saclay, Orme des Merisiers,
             91191 Gif-sur-Yvette, France}               \\[-0.2cm]
  {\protect\makebox[5in]{\quad}}  
  \\
}

\maketitle
\thispagestyle{empty}   

\begin{abstract}

We study the Potts model (defined geometrically in the cluster picture)
on finite two-dimensional lattices of size $L \times N$, with boundary
conditions that are free in the $L$-direction and periodic in the
$N$-direction. The decomposition of the partition function in terms of
the characters $K_{1+2l}$ (with $l=0,1,\ldots,L$) has previously been
studied using various approaches (quantum groups, combinatorics, transfer
matrices). We first show that the $K_{1+2l}$ thus defined actually
coincide, and can be written as traces of suitable transfer matrices in
the cluster picture. We then proceed to similarly decompose constrained
partition functions in which exactly $j$ clusters are non-contractible
with respect to the periodic lattice direction, and a partition function
with fixed transverse boundary conditions.

\end{abstract}

\section{Introduction}

The $Q$-state Potts model on a graph $G=(V,E)$ is defined initially
for $Q$ integer by the partition function
\begin{equation}
 Z=\sum_{\{\sigma\}} \exp\left[J \sum_{(i,j) \in E}
 \delta(\sigma_{i},\sigma_{j}) \right] \,,
\end{equation}
where the spins $\sigma_i=1,2,\ldots,Q$ live on the vertices $V$,
and the interaction of strength $J$ is along the edges $E$.
This definition can be extended to arbitrary real values of $Q$ through
the high-temperature expansion of $Z$, yielding \cite{FK}
\begin{equation}
 Z=\sum_{E' \subseteq E} Q^{n(E')} ({\rm e}^J-1)^{b(E')} \,, 
 \label{Zcluster}
\end{equation}
where $n(E')$ and $b(E')=|E'|$ are respectively the number of connected
components (clusters) and the cardinality (number of links) of the edge
subsets $E'$.

It is standard to introduce the temperature parameters $v={\rm e}^J-1$
and $x=Q^{-1/2}v$, and to parametrize the interval $Q\in[0,4)$ by
$Q^{1/2}=2\cos(\pi/p)=q+q^{-1}$ with $p \geq 2$ and $q=\exp(i \pi/p)$.

In two dimensions, much knowledge about the continuum-limit behaviour of the
Potts model has accumulated over the years, thanks mainly to the
progress made in conformal field theory and the theory of integrable systems.
This is particularly true at the ferromagnetic critical point, whereas much
work remains to be done in the more difficult antiferromagnetic regime.

In this paper, we shall take a different point of view, and consider a number
of combinatorial results which hold exactly true on arbitrary regular lattices
of any finite size $L \times N$, and at any temperature $x$. The choice of
boundary conditions is clearly important. In the following we shall consider
the {\em cyclic} case (free boundary conditions in the $L$-direction and
periodic in the $N$-direction), and relegate the more complicated {\em
toroidal} case (periodic boundary conditions in both directions) to a
companion paper \cite{toroidal}.

For simplicity we denote henceforth $V$ the number of vertices, $E$ the total
number of edges, and $F$ the number of faces, including the exterior one.
Also, we often consider the lattice as being built up by a transfer matrix
${\rm T}$ propagating in the $N$-direction, which we represent as horizontal.

The case of cyclic boundary conditions has already been considered by Pasquier
and Saleur \cite{PS}, where it was shown how to decompose $Z$ as a linear
combination of characters $K_{1,2l+1}$ (with $l=0,1,\ldots,L$) of
representations of the quantum group $U_q(sl(2))$. Further developments were
made independently in \cite{CS,JS}. Chang and Shrock \cite{CS} recovered
the same decomposition, but with $K_{1,2l+1}$ defined as a partial trace of
the transfer matrix ${\rm T}_{\rm spin}$ in the spin representation. Jacobsen
and Salas \cite{JS} used a similar decomposition, but with $K_{1,2l+1}$
defined as a matrix element of a transfer matrix in the cluster representation
involving two time-slices. We show here that all three points of view are
in fact equivalent, and that the characters $K_{1,2l+1}$ obtained are
identical.

Apart from that, the main part of our discussion is in the cluster picture,
following \cite{JS}. We recall the relevant definitions in
section~\ref{sec:cluster}.

The cluster configurations contributing to $K_{1,2l+1}$ turn out to be those
in which $j \ge l$ clusters are non-contractible with respect to the periodic
lattice direction. We henceforth refer to such clusters as {\em non-trivial
clusters}, or NTC for brevity. In section~\ref{sec:NTC} we give the character
decomposition of constrained partition functions $Z_{2j+1}$ in which the
number of NTC is precisely $j$. This gives as a by-product the character
decomposition of the full partition function $Z$, in agreement with
\cite{PS,CS}.

Finally, we obtain in section~\ref{sec:fixed} the character decomposition
of a partition function with {\em fixed} (rather than free) transverse
boundary conditions. The physical implications of our results are
discussed in section~\ref{sec:conclusion}.

\section{Cluster representation of the Potts model}
\label{sec:cluster}

\subsection{Transfer matrix in the cluster representation}

The cluster representation of the Potts model is defined by
Eq.~(\ref{Zcluster}). Since the clusters are non-local objects, it is not a
priori obvious how to build the partition function using a transfer matrix.
The key to tackle the problem of non-locality is to introduce a basis of
states that takes into account connectivity information \cite{BN1982}.
However, the periodic boundary conditions in the longitudinal direction
introduces a further complication, whose resolution necessitates to
introduce a transfer matrix that acts between {\em two} time slices
\cite{JS}.

\begin{figure}
  \centering
  \includegraphics[width=300pt]{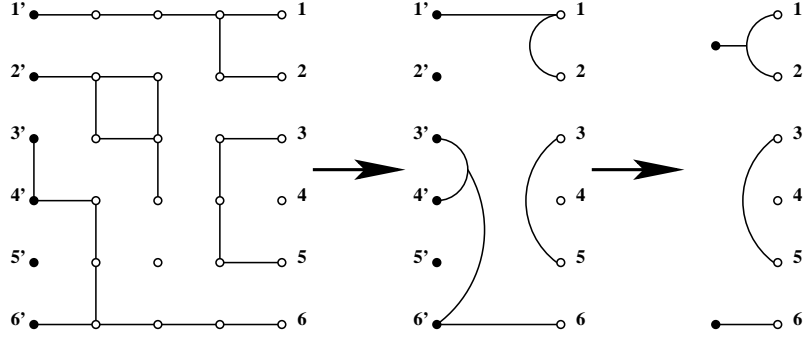}
  \caption{Example of a cluster configuration on a part of the square
 lattice with width $L=6$ (left part) and the corresponding
 connectivity state involving two time slices (middle part). The
 points in the right (resp.\ left) time slice are represented as white
 (resp.\ black) circles and are labelled $1,2,\ldots,L$ (resp.\
 $1',2',\ldots,L'$). The corresponding partition is $|v_P\rangle =
 (1'12)(2')(3'4'6'6)(5')(35)(4)$. There are two bridges, i.e.,
 independent connections between the left and right time slices. With
 the number of bridges given, the transfer matrix elements are
 independent of the connectivity information on the left time slice.
 This fact can be expressed graphically by assigning to each bridge an
 unlabelled black point and depicting the right time slice only (right
 part of the figure).}
  \label{fig1}
\end{figure}

We therefore begin by reviewing how to write the transfer matrix ${\rm T}$ in
the cluster representation when the boundary conditions are cyclic \cite{JS}.
The relevant geometry is shown in the left part of Fig.~\ref{fig1}.%
\footnote{Here, and in all subsequent figures, the explicit examples of
 configurations are for the geometry of the square lattice. We however stress
 that our reasoning is quite general and applies to an arbitrary lattice which
 is weakly regular, in the sense that the number of points in each time slice
 is equal to $L$.} Unlike the case of free boundary conditions in the
 longitudinal direction, one must take care not only of the connectivities
 inside the right time slice (at time $t=t_0$), i.e., between the points
 labelled $\{1,2,\ldots,L\}$, but also of the connectivities of the left time
 slice (at time $t=0$), i.e., between the points $\{1',2',\ldots,L'\}$, and of
 the connectivities linking the two time slices. The transfer matrix
 propagates the right time slice from time $t_0$ to time $t_0+1$. Therefore,
 the space on which the transfer matrix acts is the space of connectivity
 patterns $|v_{P}\rangle$ associated to partitions of the set
 $\{1',\ldots,L',L,\ldots,1\}$. Because of the planarity of the lattice only
 non-crossing partitions are allowed. An example of an allowed partition and
 its graphical representation is shown in the middle part of Fig.~\ref{fig1}.

A formal expression of the transfer matrix is given in \cite{JS}. Here we just
give the practical rules to calculate its elements. As in the case of free
longitudinal boundary conditions, there is a weight $v$ per coloured link and
a weight $Q$ per cluster [see Eq.~(\ref{Zcluster})], except for the clusters
containing a black circle which have a weight equal to $1$. Of particular
interest are the components of a partition that contain both white and black
circles. Such components are called bridges; we denote by $l$ the total number
of bridges in the partition (in Fig.~\ref{fig1}, $l=2$). When at a time
$t$, i.e., after applying $t$ times the transfer matrix, one obtains a state
with $l$ bridges, it means that there are $l$ clusters which begin at $t=0$
and end at a time $\ge t$. Note that the initial connectivity (at $t=0$)
is the unique state with $L$ bridges, meaning that the left and right time
slices coincide. Denoting this state $|v_L \rangle$, the
partition function $Z$ is given by
\begin{equation}
 Z= \langle u|{\rm T}^{N}|v_{L} \rangle \,,
 \label{defZ}
\end{equation}
where $\langle u |$ takes into account the periodic longitudinal boundary
conditions, by re-identifying the left and right time slices at time $t=N$ and
assigning a weight $Q$ to each of the resulting clusters \cite{JS}.

Two important observations must be made:
\begin{enumerate}
 \item ${\rm T}$ propagates the right time slice, and so, cannot modify the
 connectivity inside the left time slice.
 \item Under the action of ${\rm T}$, the number of bridges $l$ can only
 decrease or stay constant.
\end{enumerate}
These two properties imply that the transfer matrix has a lower-triangular
block form:
\begin{equation}
   {\rm T} \;=\; \left( \begin{array}{cccc}
   {\rm T}_{L,L}   & 0           & \ldots & 0 \\
   {\rm T}_{L-1,L} & {\rm T}_{L-1,L-1} & \ldots & 0 \\
   \vdots    & \vdots      &        & \vdots\\
   {\rm T}_{0,L}   & {\rm T}_{0,L-1}   & \ldots & {\rm T}_{0,0}
                           \end{array}
                    \right)
 \label{T_blocks1}
\end{equation}
Furthermore, they also imply that each block ${\rm T}_{l,l}$ on the
diagonal of ${\rm T}$ has itself a diagonal block form:
\begin{equation}
   {\rm T}_{l,l} \; = \; \left( \begin{array}{cccc}
   {\rm T}_{l,l}^{(1)}  & 0               & \ldots & 0 \\
   0                & {\rm T}_{l,l}^{(2)} & \ldots & 0 \\
   \vdots           & \vdots          &        & \vdots\\
   0                & 0               & \ldots & {\rm T}_{l,l}^{(N_l)}
                             \end{array}
                       \right)
\label{T_blocks2}
\end{equation}
Each sub-block ${\rm T}_{l,l}^{(j)}$ is characterized by a certain left slice
connectivity and a position of the $l$ bridges. Its dimension is given by the
number of compatible right slice connectivities. In fact, the $N_l$ sub-blocks
${\rm T}_{l,l}^{(j)}$, with $1\leq j \leq N_l$, are exactly equal, as the
rules for computing their matrix elements coincide. Indeed, the $L$ white
circles of the right slice do not ``see'' the left slice connectivity and from
where the $l$ bridges emanate; only the number $l$ of bridges matters. In
particular, the dimension $n(L,l)$ of the sub-block ${\rm
T}_{l,l}^{(j)}$ is independent of $j$. Moreover, because of the symmetry
between the left and right time slices, the number of sub-blocks equals their
dimension, $N_l = n(L,l)$. It can be proved that \cite{PS,CS}:
\begin{equation}
 n(L,l)=\frac{2l+1}{L+l+1} {2L \choose L-l} =
 {2L \choose L-l} - {2L \choose L-l-1} \,.
\label{defn}
\end{equation}
Note that $n(L,0)=C_L$, the $L$'th Catalan number, which is the
dimension of the cluster transfer matrix with {\em free} longitudinal
boundary conditions. Indeed, each sub-block ${\rm T}_{0,0}^{(j)}$ is
equal to the usual single time slice cluster transfer matrix
\cite{BN1982}.%
\footnote{Note that the last part of these results differ from those given in
\cite{JS}. Namely, the authors of \cite{JS} studied the chromatic polynomial
($v=-1$), so the connectivities between neighbouring points were forbidden,
and therefore the dimension of each sub-block was smaller than $n(l,L)$ given
by Eq.~(\ref{defn}). Furthermore, in the case of a square lattice, the authors
symmetrized ${\rm T}$ with respect to a top-bottom reflection of the strip.
This not only diminishes the total dimension of the transfer matrix, but also
the number of sub-blocks. At the same time it makes the structure of ${\rm T}$
slightly more complicated. Indeed, there would then be two types of
sub-blocks, depending on whether the left slice connectivity and the position
of the $l$ bridges are symmetric or non-symmetric with respect to the
reflection. The symmetrization couples either pairs of non-symmetric
sub-blocks, or pairs of states inside a symmetric sub-block. Therefore, the
symmetric and non-symmetric sub-blocks have different dimensions, the
non-symmetric sub-blocks having the largest dimension $n(L,l)$.}

Because of the block structure of ${\rm T}$, its eigenvalues are the
union of the eigenvalues of the sub-blocks ${\rm
T}_{l,l}^{(j)}$. Therefore, the sub-blocks with given $l$ being equal,
one can obtain all the eigenvalues of ${\rm T}$ by considering only
one reference sub-block for each given number of bridges $l$
\cite{JS}. For instance, one can choose as reference sub-block the one
with no connection between black circles and with $l$ bridges
beginning at $\{1',2',\ldots,l'\}$. Alternatively, one may forget the
labelling of the left time slice altogether, and simply mark by a
black point each of the components of the right-slice connectivity
which form part of a bridge.%
\footnote{Note that this choice must respect planarity: only the unnested
connectivity components (i.e., those accessible from the far left) can be
marked by a black point.}
This latter choice is represented in the right part of Fig.~\ref{fig1}. In
the following, we denote the reference sub-block simply ${\rm T}_l$.

\subsection{Definition of the characters $K_{1,2l+1}$} 

It follows from Eq.~(\ref{defZ}) and the preceding discussion that
\begin{equation}
 Z=\sum_{l=0}^L \sum_{i=1}^{n(L,l)} c(L,l,i,x) \,
 \left[ \lambda_{l,i}(L,x) \right]^N \,,
\end{equation}
where a priori the amplitudes $c$ of the eigenvalues $\lambda_{l,i}(L,x)$ ($i$
labels the distinct eigenvalues within the sub-block ${\rm T}_l$) depend of
the width $L$, the number of bridges $l$, the label $i$, and the temperature
$x$. In fact, it has been proved in \cite{PS,CS}, and used in \cite{JS}, that
$c$ depend only of $l$ (and the value of $Q$ chosen). We therefore denote them
$c^{(l)}$ in the following. Thus,
\begin{equation}
Z=\sum_{l=0}^L c^{(l)} K_{1,2l+1}(L,N,x) \,,
\label{eqfond}
\end{equation}
where the $K_{1,2l+1}(L,N,x)$ are defined as
\begin{equation}
 K_{1,2l+1}(L,N,x)=\sum_{i=1}^{n(L,l)} \left[ \lambda_{l,i}(L,x) \right]^N \,.
\label{defK}
\end{equation}
$K_{1,2l+1}$ is thus simply equal to ${\rm Tr}({\rm T}_l)^N$.

The notation $K_{1,2l+1}$ (instead of just $K_l$) is motivated by the
fact that at the ferromagnetic critical point ($x_{\rm c}=1$ for the
square lattice), and in the continuum limit, these quantities become
special cases of a generic character $K_{r,s}$ of conformal field
theory (CFT) \cite{PS}. More precisely, the character $K_{r,s}$
corresponds to the holomorphic dimension $h_{1,2l+1}$ of the CFT with
central charge $c=1-\frac{6}{p(p-1)}$. For generic (irrational) values
of $p$ this CFT is non-unitary and non-minimal. We shall comment on
the case of $p$ integer later, in section~\ref{sec:p_integer}. We
stress that we have here defined $K_{1,2l+1}$ combinatorially for an
$L \times N$ system, at any temperature $x$, with no continuum limit
being taken; we shall nevertheless refer to them as characters.

The amplitudes $c^{(l)}$ appearing in Eq.~(\ref{eqfond}) are $q$-deformed
numbers \cite{PS,CS}
\begin{equation}
 c^{(l)}=(2l+1)_q=\frac{\sin(\pi(2l+1)/p)}{\sin(\pi/p)}=
 \sum_{j=0}^l (-1)^{l-j} { l+j \choose l-j} Q^j \,.
\label{defc}
\end{equation}
Note that $c^{(l)}$ is a polynomial of degree $l$ in Q. In the next section,
we obtain a new proof of Eq.~(\ref{defc}), as a by-product of a more general
result in which we give a combinatorial sense to each term in the polynomial
separately.

In the remainder of the article, we shall decompose various partition
functions as linear combinations of the characters $K_{1,2l+1}$. Indeed, the
$K_{1,2l+1}$ are simply related to the eigenvalues of the transfer matrix and
can be considered as the basis building blocks of various restricted partition
functions.

\subsection{Equivalence with Chang and Shrock}

We now show that the $K_{1,2l+1}$, that we have defined above following
\cite{JS}, coincide with the partial traces defined in \cite{CS}.

In \cite{CS}, Chang and Shrock considered the Potts model partition function
in the spin representation: writing $Z={\rm Tr}({\rm T_{spin}})^N$ they
decomposed the spin space as a direct sum of what they called level $l$
subspaces. By definition, the level $l$ subspace corresponds to the space
generated by applying ${\rm T_{spin}}$ to the sum of spin states with $l$
spins fixed to $l$ given values. The restriction of ${\rm T_{spin}}$ to the
level $l$ subspace is exactly equal to our matrix $T_l$ (with $l$ connectivity
components marked by black points), as they have the same calculation rules
(marking a cluster with a black point corresponds to fixing its spin state,
i.e., to giving it a weight $1$ instead of $Q$) and a very similar graphical
representation of the states (resembling the right part of Fig.~\ref{fig1}).
The character $K_{1,2l+1}$ appears therefore in \cite{JS} as the restriction
of the trace to the level $l$ subspace.

We remark that the physical interpretation of the amplitudes $c^{(l)}$ made in
\cite{CS} is somewhat different from ours. Indeed, at level $l$ Chang and
Shrock considered all the independent possibilities of attributing values to
$l$ fixed spins, taking into account that some of those possibilities were
already present at lower levels. Accordingly, they interpreted $c^{(l)}$ as
the number of level $l$ states independent among themselves, and independent
of states at lower levels, and computed $c^{(l)}$ diagramatically.

Proving the equivalence of our $K_{1,2l+1}$ with those of Pasquier and Saleur
requires some further background material, and is deferred to
section~\ref{sec:eqPS}.

\section{Partition function with a fixed number of non-trivial clusters}
\label{sec:NTC}

In this section we study the character decomposition of constrained partition
functions $Z_{2j+1}$ in which the number of non-trivial clusters (NTC) is
fixed to $j$, for $j=0,1,\ldots,L$. It is important to notice that this is
different from the characters $K_{1,2l+1}$, which are related to blocks of the
transfer matrix with $l$ bridges.%
\footnote{To avoid confusion, $j$ will from now on always denote the number of
NTC in $Z_{2j+1}$, and $l$ will denote the number of bridges in $K_{1,2l+1}$.}
When imposing the periodic longitudinal boundary conditions, each bridge
becomes essentially a {\em marked} NTC. Since $K_{1,2l+1}$ may contain
further NTC which are not marked, we expect $K_{1,2l+1}$ to be a linear
combination of several $Z_{2j+1}$ with $j \ge l$. Conversely, since upon
acting with the transfer matrix the number of bridges can only decrease or
stay constant, we also expect $Z_{2j+1}$ to be a linear combination of
several $K_{1,2l+1}$ with $l \ge j$.

The primary goal of this section is to obtain the character decomposition of
$Z_{2j+1}$. In the following two subsections we therefore first express the
$K_{1,2l+1}$ in terms of the $Z_{2j+1}$, and then invert the resulting
relations.

\subsection{$K_{1,2l+1}$ in terms of $Z_{2j+1}$}

Recalling that $K_{1,2l+1}={\rm Tr}\, ({\rm T}_{l})^{N}$, we can write 
\begin{equation}
  K_{1,2l+1}=\sum_{i=1}^{n(L,l)} \langle v_{l,i} | {\rm T}^N |v_{l,i} \rangle
  \,,
\label{exprtr}
\end{equation}
where the $|v_{l,i} \rangle$ are the $n(L,l)$ possible connectivity
states with $l$ bridges, i.e., states such as those shown in the right
part of Fig.~\ref{fig1} with $l$ black points.

\begin{figure}
  \centering
  \includegraphics[width=350pt]{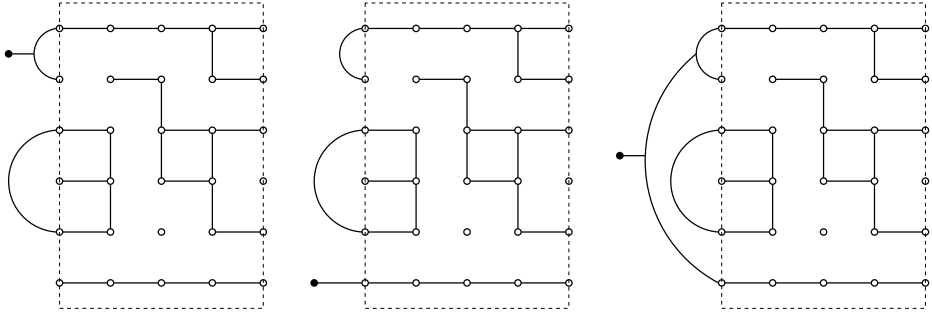}
  \caption{A cluster configuration on a portion of the square lattice
    (shown inside a dashed box for clarity) and the three compatible
    connectivity states (shown on the left of each copy of the cluster
    configuration). In each of the three cases, the final connectivity
    (i.e., the way in which the $L$ points on the rightmost column of
    the lattice are interconnected and marked by black points through
    the cluster configuration {\em and} the connectivity state on the
    left) is equal to the initial connectivity state.}
  \label{fig3}
\end{figure}

We first show that a given cluster configuration with $j$ NTC is contained
$n(j,l)$ times in $K_{1,2l+1}$. To this end, we define that a connectivity
state $|v_{l,i} \rangle$ is {\em compatible} with a given cluster
configuration if the action of the cluster configuration on $|v_{l,i} \rangle$
(in the sense of a transfer matrix acting towards the right) yields the same
connectivity $|v_{l,i} \rangle$. An example is shown in Fig.~\ref{fig3}. It is
useful to ``forget'' for a moment that the longitudial boundary conditions are
cyclic, i.e., to consider the leftmost and rightmost columns of the lattice as
distinct. Indeed, the periodic boundary conditions are already encoded in the
fact that the final and initial states in Eq.~(\ref{exprtr}) must coincide.
The goal is then to show that any cluster configuration with $j$ NTC is
compatible with precisely $n(j,l)$ different connectivity states.

Consider then a given cluster configuration with $j$ NTC, with the
$k$'th NTC ($k=1,2,\ldots,j$) connecting onto the points $\{y_k\}$ of the
rightmost column. For example, in Fig.~\ref{fig3} we have
$\{y_1\}=\{1,2\}$ and $\{y_2\}=\{6\}$. The connectivity states
$|v_{l,i} \rangle$ compatible with the cluster configuration
can be constructed as follows:
\begin{enumerate}
\item The connectivities of the points $y \notin \cup_{k=1}^j \{y_k\}$
  must be connected in the same way in $|v_{l,i} \rangle$ as in the
  cluster configuration. For instance,
  in Fig.~\ref{fig3} the points $y=3,5$ must be connected.
\item The points $\{y_k\}$ within the same bridge (for example, $y=1,2$
  in Fig.~\ref{fig3}) must be connected in $|v_{l,i} \rangle$.
\item One can independently choose to associate or not a black point to each
  of the sets $\{y_k\}$. One is free to connect or not two distinct sets
  $\{y_k\}$ and $\{y_{k'}\}$.
\end{enumerate}

Clearly, the rules 1 and 2 leave no choice. The rule 3 implies in particular
that $j \ge l$, or else there is no compatible state $|v_{l,i} \rangle$.
The choices mentioned in rule 3 then leave us $n(j,l)$ possibilities for
constructing a compatible $|v_{l,i} \rangle$.

We have therefore shown that a given cluster configuration with $j$
NTC is contained $n(j,l)$ times in $K_{1,2l+1}$. As $K_{1,2l+1}$ is
simply a trace, each of its NTC carries a weight of 1, whereas the $j$
NTC in $Z_{2j+1}$ each have the usual cluster weight of $Q$. We
therefore arrive at the result
\begin{equation}
K_{1,2l+1}=\sum_{j=l}^L n(j,l) \frac{Z_{2j+1}}{Q^j}
\label{exprK}
\end{equation}
where we recall that $n(j,l)$ has been defined in Eq.~(\ref{defn}). 

\subsection{$Z_{2j+1}$ in terms of $K_{1,2l+1}$}

Inverting the relations (\ref{exprK}) yields
\begin{equation}
 Z_{2j+1}=\sum_{l=j}^L c_j^{(l)} K_{1,2l+1}
\label{exprZ}
\end{equation}
with the coefficients $c_j^{(l)}$ given by
\begin{equation}
 c_j^{(l)}=(-1)^{l-j} {l+j \choose l-j} Q^j \,.
 \label{defg}
\end{equation}
An interesting special case, which we will refer to in the following,
is obtained for $j=0$, i.e., by disallowing any NTC. {}From
Eqs.~(\ref{exprZ})--(\ref{defg}), we obtain an alternating sum of the
$K_{1,2l+1}$:
\begin{equation}
 Z_1=\sum_{l=0}^L (-1)^l K_{1,2l+1}
 \label{exprZ1}
\end{equation}

Note also that the total partition function of the Potts model is given by
\begin{equation}
 Z=\sum_{j=0}^L Z_{2j+1} \,.
 \label{Zsum}
\end{equation}
Comparing Eqs.~(\ref{defg}) and (\ref{defc}) we infer that
\begin{equation}
c^{(l)}=\sum_{j=0}^l c_j^{(l)}
\end{equation}
and from Eqs.~(\ref{exprZ}) and (\ref{Zsum}) we obtain as promised
Eq.~(\ref{eqfond}) for the full partition function.

Interestingly, then, the effect of fixing the number of NTC to $j$
is to keep only the term multiplying $Q^j$ in the expression (\ref{defc}) of
$c^{(l)}$. As $c^{(l)}$ is polynomial of degree $l$ in $Q$, only the
$K_{1,2l+1}$ with $l\geq j$ contribute to the character decomposition of
$Z_{2j+1}$. This is in agreement with the physical argument given at
the beginning of section~\ref{sec:NTC}.

\subsection{Equivalence with Pasquier and Saleur}
\label{sec:eqPS}

We can now prove that the $K_{1,2l+1}$ defined in \cite{PS} using the
six-vertex model are equal to the $K_{1,2l+1}$ we defined in
Eq.~(\ref{defK}) using the cluster transfer matrix. Before attacking
the proof, let us briefly recall where the connection with the
six-vertex model comes from.

On a planar lattice, the cluster representation of the Potts model
partition function is equivalent to a loop representation, where the
loops are defined on the medial lattice and surround the clusters
\cite{Baxter_book}. {}From Eq.~(\ref{Zcluster}) and the Euler
relation, the weight of a loop configuration $E'$ is $Q^{(V+c(E'))/2}
x^{b(E')}$, where $c(E')$ is its number of loops.%
\footnote{Note that we do not factorize $Q^{V/2}$, in order to recover
exactly the same expression for the $K_{1,2l+1}$ as before.}  An
oriented loop representation is obtained by independently assigning an
orientation to each loop, with weight $q$ (resp. $q^{-1}$) for
counterclockwise (resp. clockwise) loops (recall that
$Q^{1/2}=q+q^{-1}$). In this representation one can define the spin
$S_z$ along the transfer direction (with parallel/antiparallel loops
contributing $\pm 1/2$) which acts as a conserved quantum number. Note
that $S_z=l$ means that there are at least $l$ non-contractible loops,
i.e., loops that wind around the periodic ($N$) direction of the
lattice.  Indeed, the contractible loops do not contribute to
$S_z$.

The weights $q^{\pm 1}$ can be further redistributed locally, as a
factor of $q^{\alpha /2\pi}$ for a counterclockwise turn through an
angle $\alpha$ \cite{Baxter_book}. While this redistribution correctly
weighs contractible loops, the non-contractible loops are given weight
$2$, but this can be corrected \cite{PS} by twisting the model, i.e.,
by inserting the operator $q^{2S_z}$ into the trace that defines the
partition function.  A partial resummation over the oriented-loop
splittings at vertices which are compatible with a given orientation
of the edges incident to that vertex now gives a six-vertex model
representation \cite{Baxter_book}. Each edge of the medial lattice
then carries an arrow, and these arrows are conserved at the vertices:
the net arrow flux defines $S_z$ as before. The six-vertex model again
needs twisting by the operator $q^{S_z}$ to ensure the correct
weighting. Considering each arrow as a spin $1/2$, the transfer matrix
in the six-vertex representation, ${\rm T_{6V}}$, acts on a quantum
chain of $2L$ spins $1/2$. ${\rm T_{6V}}$ can be expressed in terms of
generators of a Temperley-Lieb algebra, and therefore commutes with
the generators of the quantum group $U_q(sl(2))$ \cite{PS}. In
addition to $S_z$ one can then define the total spin $S$
(corresponding to the Casimir).

In this subsection we now follow \cite{PS} and define $K_{1,2l+1}$ as
the trace of $({\rm T_{6V}})^N$ in the space of highest weights of spin
$S=S_z=l$.%
\footnote{Note that in this context, Eq.~(\ref{defc}) follows by noting that
each irreducible representation contains $2l+1$ states, which is replaced by
the $q$-deformed number $(2l+1)_q$ on account of the twist.}
With this definition, our goal is to decompose $K_{1,2l+1}$ in terms of the
$Z_{2j+1}$, obtaining again Eq.~(\ref{exprK}), from which we shall conclude
that the two definitions of $K_{1,2l+1}$ are equivalent.

To this end, we first remark that
\begin{equation}
 K_{1,2l+1}=F_{2l+1}-F_{2(l+1)+1} \,,
\label{exprKF}
\end{equation}
where $F_{2l+1}$ is the trace of $({\rm T_{6V}})^N$ on the space of {\em all}
states of spin $S_z=l$. Indeed, the number of highest weight states of spin
$S=S_z=l$ equals the number of states of spin $S_z=l$ minus the number of
states of spin $S_z=l+1$. Therefore, we first decompose $F_{2l+1}$. The
advantage of working with $F_{2l+1}$ is that only $S_z$ is specified, not $S$.
Indeed, only $S_z$ has a simple interpretation in the oriented loop
representation: a basis of the space corresponding to $S_z=l$ is given
simply by all states with a net arrow flux of $l$ to the right, whereas the
states with $S=S_z=l$ would be more complicated linear combinations of given
spin configurations.

\begin{figure}
  \centering
  \includegraphics[width=150pt]{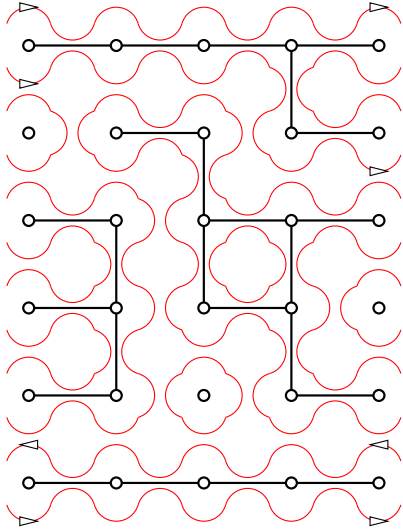}
  \caption{Loop configuration corresponding to the cluster
 configuration in Fig.~\ref{fig3}. The contractible loops can have any
 orientation (not shown), whereas those of the non-contractible loops
 are constrained by the chosen value of $S_z$. With $2j=4$
 non-contractible loops we show one of the four possible orientations
 leading to $S_z=1$.}
  \label{fig4}
\end{figure}

We now consider a configuration of oriented loops contributing to
$Z_{2j+1}$, i.e., with $2j$ non-contractible loops. As the
contractible loops do not contribute to $S_z$, there are no
constraints on their orientations.  Among the $2j$ non-contractible
loops, $j+l$ (resp.\ $j-l$) must be oriented to the right (resp.\
left) in order to obtain $S_z=l$ (recall that $l \le j$). This is
illustrated in Fig.~\ref{fig4}. There are therefore ${2j \choose j-l}$
possible orientations of the non-contractible loops compatible with
the chosen value of $S_z$. Correcting for the factors of $Q$ as
before, we conclude that the character decomposition of $F_{2l+1}$ is
\begin{equation}
  F_{2l+1}=\sum_{j=l}^L {2j \choose j-l} \frac{Z_{2j+1}}{Q^j} \,.
\label{exprFl}
\end{equation}
Using now Eq.~(\ref{exprKF}), and keeping in mind the identity in
Eq.~(\ref{defn}), we finally obtain Eq.~(\ref{exprK}). This proves
that our definition of $K_{1,2l+1}$ coincides with the one used in
\cite{PS}.

\subsection{Case of $p$ integer}
\label{sec:p_integer}

When $p$ is integer, $U_{q}(sl(2))$ mixes representations
with $l'=p-1-l+ n p$ and $l'=l+ n p$, with $n$ integer. Of
particular interest are the type II representations, and it
can be shown that the traces on highest weight states of type II
are given by \cite{PS}
\begin{equation}
 \chi_{1,2l+1}(L,N,x) \;=\;
 \sum_{n \ge 0} \left( K_{1,2(np+l)+1}(L,N,x) -
                       K_{1,2((n+1)p-1-l)+1}(L,N,x) \right) \,.
\label{defchi}
\end{equation}
For convenience in writing Eq.~(\ref{defchi}) we have defined
$K_{1,2l+1}(L,N,x) \equiv 0$ for $l > L$. At the ferromagnetic
critical point, and in the continuum limit, the quantities
$\chi_{1,2l+1}$ become characters corresponding to primary fields of
the unitary, minimal model $M_{p,p-1}$ with central charge
$c=1-\frac{6}{p(p-1)}$. The many cancellations in Eq.~(\ref{defchi})
are linked to the existence of null vectors in the corresponding
irreducible Verma modules.  In fact, Eq.~(\ref{defchi}) is then
nothing else than the Rocha-Caridi equation \cite{Rocha-Caridi}.

As in the case of the generic characters $K_{1,2l+1}$, the definition
(\ref{defchi}) of the minimal characters $\chi_{1,2l+1}$ is at finite
size, and for any temperature $x$, but by analogy we shall still refer
to $\chi_{1,2l+1}(L,N,x)$ as a minimal character.

It does not appear to be possible to compute the $\chi_{1,2l+1}$
directly in the cluster representation, i.e., otherwise than by first
computing the corresponding $K_{1,2l'+1}$ and then applying
Eq.~(\ref{defchi}). They can however be computed directly in an
$A_{p-1}$ type RSOS model \cite{Pasquier_87} with specific boundary
conditions \cite{SB}.

Many, but not all, character decompositions of partition functions in
terms of $K_{1,2l+1}$ turn into character decompositions in terms of
$\chi_{1,2l+1}$ for $p$ integer. This is the case for the total
partition function, due to the symmetries
\begin{equation}
c^{(l)}=-c^{(p-1+np-l)}=c^{(np+l)} \,.
\label{symc}
\end{equation}
Therefore, using Eq.~(\ref{eqfond}), one obtains \cite{SB}
\begin{equation}
 Z=\sum_{l=0}^{\lfloor (p-2)/2 \rfloor} c^{(l)} \chi_{1,2l+1} \,.
\end{equation}
Note that the sum contains less terms than
before; in fact it is over those minimal characters that would be inside
the Kac table at the ferromagnetic critical point \cite{Yellow_book}.

On the other hand, the formula for the $Z_{2j+1}$, when the number of
NTC is fixed to $j$, cannot in general be expressed in terms of
the $\chi_{1,2l+1}$ for $p$ integer. One interesting exception is for
$j=0$ (no NTC allowed) and $p$ even. Using Eq.~(\ref{exprZ1}) one
obtains
\begin{equation}
  Z_1=\sum_{l=0}^{\lfloor (p-2)/2 \rfloor} (-1)^l \chi_{1,2l+1} 
  \qquad \mbox{($p$ even)}\,.
\end{equation}
This effect of parity in $p$ is present in many other properties of the
RSOS models \cite{RJ2}.

\section{Fixed transverse boundary conditions}
\label{sec:fixed}

Another constrained partition function whose character decomposition
would be of interest is that of the Potts model on a cyclic lattice
strip with {\em fixed} boundary conditions on the upper and lower
horizontal row of Potts spins. It turns out to be easier to obtain the
decomposition of a slightly modified object, namely the corresponding
partition function on the dual lattice, with fixed boundary conditions
on the two dual spins each of which lives on an exterior infinite
face.

\subsection{A modified model on the dual lattice}

We consider therefore $\tilde{Z}_{Q_0}(\tilde{x})$, the partition
function of the Potts model, defined on the lattice dual to the $L
\times N$ cyclic strip considered in the preceding sections, evaluated
at the dual temperature $\tilde{x}=1/x$. For the sake of
generality, any dual cluster which contains one (or both) exterior dual
vertices has a weight of $Q_0$ instead of $Q$. Note that $Q_0=1$
corresponds to fixed boundary conditions on the two exterior dual
spins. The case $Q_0=Q$ is equivalent (under duality) to the free
transverse boundary conditions considered above; we denote the
corresponding dual partition function $\tilde{Z}(\tilde{x})$.

\begin{figure}
  \centering
  \includegraphics[width=150pt]{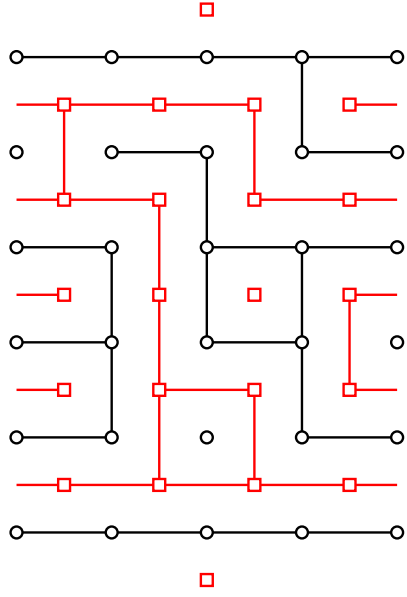}
  \caption{Direct and dual clusters corresponding to the configuration
  in Fig.~\ref{fig4}. Direct (resp.\ dual) vertices are shown as black
  circles (resp.\ red squares). There are two direct NTC and three dual
  NTC (see text).}
  \label{fig5}
\end{figure}

We search the character decomposition of $\frac{Q^{2-F}
v^E}{Q_0}\tilde{Z}_{Q_0}(\tilde{x})$, where the prefactor is chosen so
as to make the final result simpler.  To achieve this goal, one needs
first to convert the weights of the dual clusters into weights of
direct clusters. Indeed, by duality a direct cluster configuration is
in one-to-one correspondence with a dual cluster configuration
\cite{Baxter_book}, as shown in Fig.~\ref{fig5}. To simplify the
notation, we adopt the following convention: a dual cluster is called
a {\em non-trivial cluster} (NTC) if it is non-contractible with
respect to the periodic lattice direction, or if it contains one (or
both) of the exterior dual spins.  With this convention, a dual
configuration with $j+1$ dual NTC corresponds always to a direct
configuration with $j$ direct NTC. Note that there is always at least
one dual NTC.

Given a cluster configuration, we denote by $t$ the number of direct
trivial (contractible) clusters, by $\tilde{t}$ the number of dual
trivial clusters, by $b$ the number of direct edges, and by
$\tilde{b}$ the number of dual edges. Consider now the weight of a
configuration with $j+1$ dual NTC in $\frac{Q^{2-F}
v^E}{Q_0}\tilde{Z}_{Q_0}(\tilde{x})$. For $j \ge 1$ (resp.\ $j=0$)
this is $\frac{Q^{2-F} v^E}{Q_0} Q_0^2 Q^{j-1} Q^{\tilde{t}}
\tilde{v}^{\tilde{b}}$ (resp.\ $\frac{Q^{2-F} v^E}{Q_0} Q_0 Q^{\tilde{t}}
\tilde{v}^{\tilde{b}}$), since the two exterior dual vertices are
contained in two different (resp.\ the same) dual NTC. We have here
denoted the dual parameter $\tilde{v}=Q/v$.

To express these weights in terms of the direct quantities, we recall
the fundamental duality relation \cite{Baxter_book} $Q^{1-F} v^E
\tilde{Z}(\tilde{x})=Z(x)$, valid because the lattice is planar.
Translated into a relation on the weights of a single cluster configuration
this reads
\begin{equation}
 Q^{1-F} v^E Q^{j+1} Q^{\tilde{t}} \tilde{v}^{\tilde{b}}= Q^j Q^t v^b \,.
\label{dual}
\end{equation}
Therefore, the weight of a cluster configuration with $j$ direct NTC
reads $Q_0 Q^{j-1} Q^t v^b$ if $j\geq 1$, and $Q^t v^b$ if $j=0$. We
thus deduce the following result: the weight of a direct cluster
configuration in $\frac{Q^{2-F} v^E}{Q_0}\tilde{Z}_{Q_0}(\tilde{x})$ is
the same as in $Z(x)$, except that for $j \ge 1$ direct NTC, one of the
NTC has a weight $Q_0$ instead of $Q$.

\subsection{$\tilde{Z}_{Q_0}(\tilde{x})$ in terms of $K_{1,2l+1}$}

Let us recall that when inserting the development (\ref{defc}) of
$c^{(l)}$ into Eq.~(\ref{eqfond}) for $Z$, we have a geometrical
interpretation for each term separately: from Eq.~(\ref{exprZ}) the
term in $Q^j$ gives precisely $Z_{2j+1}$. Due to the result given
after Eq.~(\ref{dual}), we must now simply replace $Q^{j}$ by $Q_0
Q^{j-1}$ for $j\geq 1$ and keep unchanged the term corresponding to
$j=0$. Therefore
\begin{equation}
 \frac{Q^{2-F} v^E}{Q_0}\tilde{Z}_{Q_0}(\tilde{x})=
 \sum_{l=0}^L b^{(l)} K_{1,2l+1}(x)
 \label{exprZdu}
\end{equation}
with the amplitudes
\begin{equation}
 b^{(l)}= \frac{Q_0}{Q} c^{(l)} + (-1)^l \left( 1-\frac{Q_0}{Q} \right)
        = (-1)^l +\sum_{j=1}^l (-1)^{l-j} {l+j \choose l-j} Q_0 Q^{j-1} \,.
\label{defb}
\end{equation}
Note that when $Q_0=Q$, we recover $b^{(l)}=c^{(l)}$ as we should.

Just like in the case of free transverse boundary conditions, each
power of $Q$ in Eq.~(\ref{exprZdu}) can be interpreted separately as a
partition function with a fixed number of NTC.

Let us consider a couple of limiting cases of Eq.~(\ref{exprZdu}).
For $Q_0 \rightarrow 0$, $b^{(l)}=(-1)^l$ and therefore
\begin{equation}
 {\rm lim}_{Q_0 \rightarrow 0} 
 \left( \frac{Q^{2-F} v^E}{Q_0}\tilde{Z}_{Q_0}(\tilde{x}) \right) =
 \sum_{l=0}^L (-1)^l K_{1,2l+1}(x) = Z_1(x) \,,
\end{equation}
where we have used Eq.~(\ref{exprZ1}). We thus recover exactly the
partition function with no direct NTC.

On the other hand, for $Q_0 \to \infty$, there is no $K_{1,1}$ in the
expansion of $\frac{Q^{2-F} v^E}{Q_0}\tilde{Z}_{Q_0}(\tilde{x})$, i.e.,
$l=0$ is forbidden.  This is indeed expected, since in that limit
there can be no dual cluster connecting the two exterior vertices, and
therefore there is at least one direct NTC.  Thus $j=0$ is forbidden,
and since $l\geq j$, we deduce that $l=0$ is forbidden as well.

We now consider the case of $p$ integer. Using Eqs.~(\ref{defb}) and  
(\ref{symc}), we obtain that for $p$ even
\begin{equation}
b^{(l)}=-b^{(p-1+np-l)}=b^{(np+l)} \,,
\label{symb}
\end{equation}
and we can write
\begin{equation}
 \frac{Q^{2-F} v^E}{Q_0}\tilde{Z}_{Q_0}(\tilde{x})=
 \sum_{l=0}^{\lfloor (p-2)/2 \rfloor} b^{(l)} \chi_{1,2l+1}(x) 
 \qquad \mbox{($p$ even)}\,.
 \label{exprZduchi}
\end{equation}

Note finally that $b^{(1)}=Q_0-1$. This means that with fixed cyclic
boundary conditions ($Q_0=1$) the term $l=1$ drops out from the
character decomposition. This fact has been exploited in a recent
study of partition function zeroes of the RSOS models \cite{JRS}.

\subsection{Square lattice model with $Q_0=1$}

The case of $Q_0=1$ can be interpreted in the spin representation as
having the same fixed value of the dual spins on the two exterior dual
vertices.  Alternatively, in the cluster picture, a dual cluster
containing one or both exterior vertices has the weight $1$ instead of
$Q$.

Suppose now for simplicity that the direct lattice is a square
lattice. The dual lattice is then a square lattice too, except for the
two exterior vertices, each of which is equivalent to an extra line of
spins all fixed in the same state. To make the equivalence perfect we
should include an extra global factor of $\exp (2NJ)$, because of the
interactions between spins inside each of the two extra lines (see
Fig.~\ref{fig5}).  The dual lattice is thus equivalent to a square
lattice of width $L+1$ and of length $N$, with periodic boundary
conditions along $N$ and all the spins at the boundaries fixed to the
same value. We denote the corresponding partition function $Z_{\rm
ff}(L+1,N,x)$.  Eq.~(\ref{exprZdu}) then reads explicitly
\begin{equation}
 Z_{\rm ff}(L,N,x)=
 \frac{\exp (2NJ)}{Q^{2-F} v^E}
 \sum_{l=0}^L b^{(l)} K_{1,2l+1}(L-1,N,\tilde{x}) \,.
\label{Z++}
\end{equation}

Let us write out the explicit results for integer $Q$.
For the Ising model ($Q=2$ or $p=4$) we have
\begin{equation}
  Z_{\rm ff}(L,N,x)=
  \frac{\exp (2NJ)}{2^{2-F} v^E} \chi_{1,1}(L-1,N,\tilde{x}) \,,
  \label{Z++2}
\end{equation}
while for the three-state Potts model ($Q=3$ or $p=6$) we find
\begin{equation}
  Z_{\rm ff}(L,N,x)=
  \frac{\exp (2NJ)}{3^{2-F} v^E} (\chi_{1,1}(L-1,N,\tilde{x})+
  \chi_{1,5}(L-1,N,\tilde{x}))
\label{Z++3}
\end{equation}
In the latter case, it is interesting to note that at the
ferromagnetic critical point $\chi_{1,1}+\chi_{1,5}$ is nothing but
the character of the identity operator with respect to the extended
$W_3$ algebra \cite{Cardy89}.

\section{Conclusion}
\label{sec:conclusion}

We have explained in this paper how to decompose various constrained
partition functions of the Potts model with cyclic boundary conditions
in terms of the characters $K_{1,2l+1}$. These decompositions, whose
origin is purely combinatorial, hold true in finite size, for any
weakly regular lattice, and at any temperature $x$.

In particular we can decompose the ratios $Z_{2j+1}/Z$, which are the
probabilities of having exactly $j$ non-trivial clusters. While these
probabilities are well-understood in the continuum limit, at the
ferromagnetic critical point at least, our results shed more light
on their fine structure, in particular regarding corrections to
scaling.

Finally, we have seen that fixed transverse boundary conditions lead to the
disappearance of the term with $l=1$. Physically, one would expect the
breaking of the $S_Q$ permutation symmetry of the spin states induced by the
fixed boundary conditions to simplify the structure of the complex-temperature
phase diagram in the low-temperature phase. This expectation is indeed brought
out in a recent numerical study \cite{JRS}.

\vspace{0.1cm}

\noindent
{\bf Acknowledgments.}

JLJ thanks the members of the SPhT, where part of this work was done,
for their kind hospitality.

\end{document}